\documentclass[sigconf]{acmart}

\usepackage{url}
\usepackage{amsmath}
\usepackage{balance}
\usepackage{booktabs}
\usepackage{tabularx}
\usepackage{graphicx}

\usepackage{fancyhdr} 
\fancypagestyle{firststyle}
{
   \fancyhf{}
   \fancyfoot[C]{\scriptsize{Proceedings of the 5th International Conference on E-Society, E-Education and E-Technology (ICSET 2021), Taipei (online), ACM, pp.~157--163. This is the authors' copy. The publisher's copy is available online via \url{https://doi.org/10.1145/3485768.3485815}}}
   }

\begin{document}

\title{Digital Divides and Online Media}

\author{Jukka Ruohonen}
\affiliation{\institution{University of Turku\country{Finland}}}
\email{juanruo@utu.fi}

\author{Anne-Marie Tuikka}
\affiliation{\institution{University of Turku\country{Finland}}}
\email{anne-marie.tuikka@utu.fi}


\begin{abstract}
Digital divide has been a common concern during the past two or three decades;
traditionally, it refers to a gap between developed and developing countries in
the adoption and use of digital technologies. Given the importance of the topic,
digital divide has been also extensively studied, although, hitherto, there is
no previous research that would have linked the concept to online media. Given
this gap in the literature, this paper evaluates the ``maturity'' of online
media in 134 countries between 2007 and 2016. Maturity is defined according to
the levels of national online media consumption, diversity of political
perspectives presented in national online media, and consensus in reporting
major political events in national online media. These aspects are explained by
considering explanatory factors related to economy, infrastructure, politics,
and administration. According to the empirical results based on a dynamic panel
data methodology, variables representing each aspects are also associated with
the maturity of national online media.
\end{abstract}

%

\ccsdesc[500]{Social and professional topics~Computing / technology policy}

\keywords{Media systems; plurality; freedom of expression; civil rights; developing countries; time-series cross-sectional; comparative research}

\maketitle

\section{Introduction}

\thispagestyle{firststyle} 

Digital divide is a well-established umbrella concept traditionally used to
describe the gap that exists in the adoption, use, and maturity of digital
technologies---from personal computers to national network
infrastructures---between developed and developing countries. But while in the
1990s the concept was mainly used in terms of access to digital
infrastructures---among these the Internet, the concept's scope has widened step
by step. Today, there are not only a digital divide, or the digital divide, but
digital divides~\cite{Norris01}; between countries, between infrastructures,
between education systems, between disabled and those without
impairments~\cite{Sachdeva15}, between young and old, between genders, and so on
and so forth.

The present paper takes a conventional cross-country viewpoint to these digital
divides. That said, the paper's context and perspective are both unconventional
and novel; the paper is the first to examine a digital divide in terms of online
media and its maturity particularly in developing countries. Furthermore,
according to a decent literature search from all relevant databases, there is a
very limited literature base on media in developing countries. Against this
background, to motivate the context and perspective further, a few words should
be said more about digital~divides and media.

On one hand, these digital divides cast a shadow on popular explanations for the
use digital technologies. For instance, the so-called technology acceptance
model (TAM) posits that people tend to accept new digital technologies when
these are easy to use and benefit their daily work~\cite{Davis89}. Although the
model has been extended with various other factors, it still tends to undermine
socioeconomic factors. When one does not know how to read, or when one is
suffering from hunger, ease of use or perceived usefulness of new technologies
are supposedly the last thing in one's mind. A similar point applies to online
media; even though audiovisual media content is increasingly common, at least a
degree literacy is still required to appreciate media in general and journalism
in particular. On the other hand, there are certain shortcomings and historical
baggage in the digital divide(s) research as well. Although access to the
Internet and digital technologies in general continues to be a grave problem in
some developing countries, as it was already in the 1990s, in some developing
countries Internet access is already on par with developed countries thanks to
widespread use of mobile phones~\cite{Karar19, Norris01, Ogondo16}. Therefore,
the infrastructure is already there for online media at least in some developing
countries. A more fundamental point is also present; it is relevant to consider
other factors than technology when considering the digital divide(s).

The reasons for not being online are many. Some may not have access, but others
may not have money, skills, time, or even need; and others may lack awareness or
interests~\citep{Vartanova12}. In other words, digital divide is also a
socioeconomic and cultural phenomenon~\cite{Ogondo16}. Access to digital devices
and the Internet do not guarantee continuous and productive use of digital
technologies and services, which, throughout the world, are increasingly
important for citizens to efficiently participate in society and
politics~\cite{Karar19}. Even in Western countries literature skills, for
instance, tend to some extent prevent efficient participation and thus undermine
citizens' self-management and self-actualization~\cite{Ruohonen21ICEDEG}. As is
soon discussed in Section~\ref{sec: framework}, media has an important function
in this conundrum.

Although economic rationale is present, media as a system is often seen to also
involve societal and educational functions; to enlighten and empower citizens;
to reduce prejudices~\citep{Lissitsa19}; to provide a public sphere on which
societal disagreements and conflicts can be discussed and mediated; and so
on. When considering the emergence of a mature online media system particularly
in developing countries, technological and economic factors cannot thus provide
a sufficient explanation alone. This point can be further used to criticize the existing technology adoption and evolution models.

Besides the classical TAM model and its adaptations, there is a large number of
different maturity models for digital businesses. However, many---if not
most---of these models concentrated on organizational and firm-level economic
factors, including transactions and business processes~\cite{Looy13,
  Morais12}. Although online media companies are businesses like other
companies, their operating environment is distinctively different from typical
e-business companies. While property rights, intellectual rights, and contract
law in general are necessary for all businesses~\cite{Williamson91a}, media
companies typically (but not necessarily) need also civil rights to succeed;
freedom of expression, freedom of speech, and media freedom are inherently tied
to media and journalism. A media company may succeed economically without these,
but the success comes at the cost of sacrificing almost universally held ideals
about journalism. These points motivate a societal aspect discussed in the next
Section~\ref{sec: framework}. Then, the empirical research design is elaborated
in Section~\ref{sec: research design}, results are presented in
Section~\ref{sec: results}, and conclusions follow in Section~\ref{sec:
  conclusion}.

\section{A Framework}\label{sec: framework}

Digital increasingly intervenes with physical. Not only are new technologies
such as artificial intelligence and virtual reality making the boundary blurry
for human beings, but the boundary spanning extend toward political economy,
geopolitics, and even warfare~\cite{Ruohonen21MIND}. Also media is entangled in
this ``mega trend'' of digitalization. As a format, traditional print media has
long lost to radio, television, and online media sources---including online-only
outlets and social media---in most Western countries particularly among the
younger generations~\cite{DNR20}. Media continues to be in a state of
rupture. But traditional media has not stood idle. Most newspapers are available
online, new online-only outlets have emerged, and innovative platforms have been
developed to enhance journalism in the new digital era. At the same time, large
multinational companies are eating media's advertising
revenues~\cite{Ruohonen21DJ}, misinformation, disinformation, and propaganda are
rampant online~\cite{Ruohonen21TFSC}, and in many countries local newspapers are
struggling~\cite{Sands19}, media ownership continues to
concentrate~\cite{Grisold96}, and public broadcasting media is under a political
pressure~\cite{Dawes14}, to only name a few challenges. These are some of the
smaller trends behind the digitalization mega trend in Western
countries. Elsewhere, in developing countries, the situation may be somewhat
different. It is difficult to face digitalization of media when there is a
shortage of electricity---to give an extreme example.

How to theorize the maturity of online media particularly in developing
countries? To keep things simple, a sensible starting point is to distinguish a
\textit{societal perspective} from an \textit{economic perspective} to online
media. To still simplify, infrastructure can be considered alongside economy,
and the societal viewpoint can be broken down to administration and
politics. The analytical result is illustrated in Fig.~\ref{fig:
  framework}. Needless to say, in reality, all of the analytical dimensions
intervene; there is no economy without an infrastructure, politics and economy
result in political economy, and so forth and so on.

\begin{figure}[th!b]
\centering
\includegraphics[width=\linewidth, height=5.5cm]{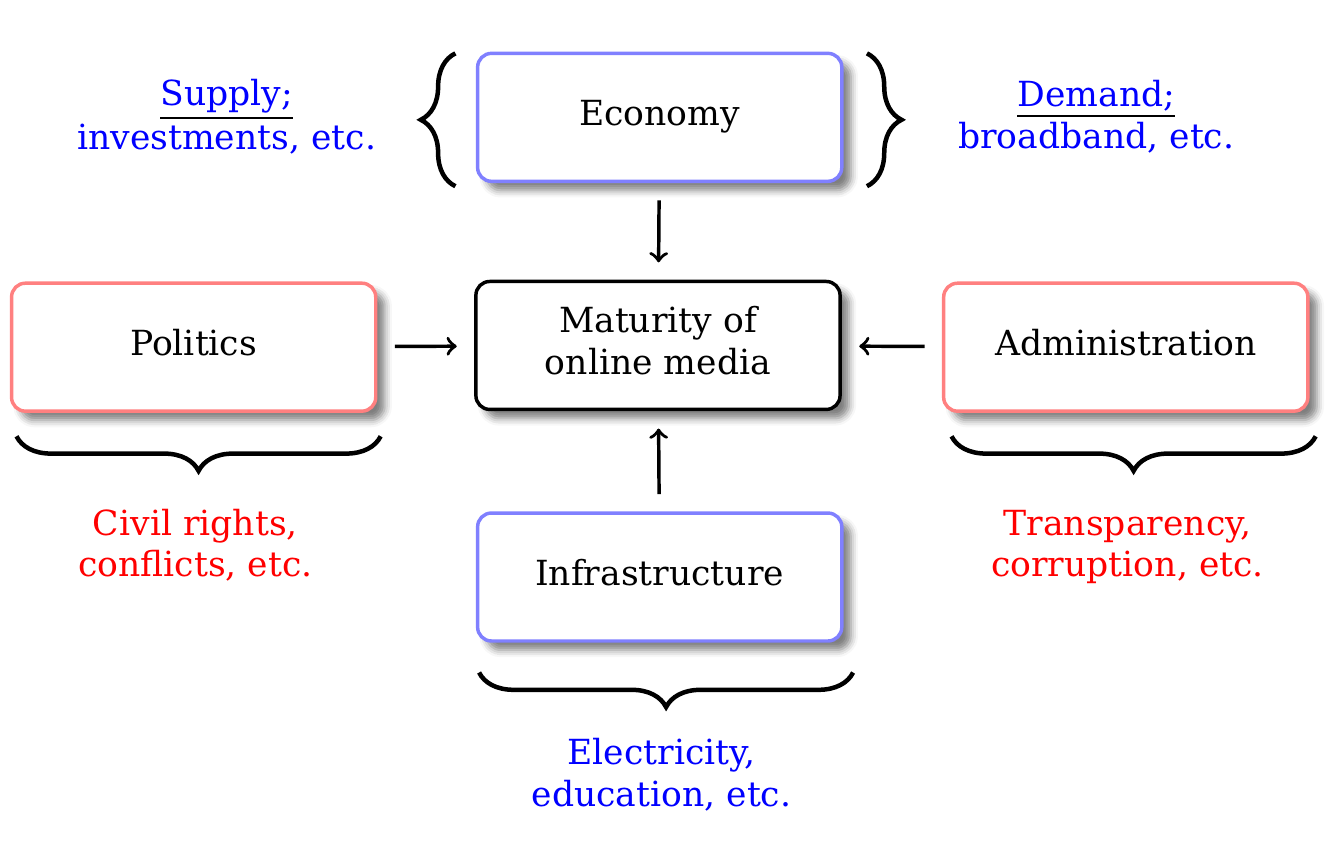}
\caption{An Analytical Framework}
\label{fig: framework}
\end{figure}

The societal perspective is most of all a normative stance on what media is and
should be. This is the question that was asked by Siebert and associates in
their classical 1984 monograph \textit{Four Theories of the
  Press}~(\cite{Siebert84}; for the background see also \cite{Clarke06,
  Ruohonen21EJC}). Although the book starts to show its age, it is still useful
for understanding the normative underpinnings. Accordingly, there are four
normative theories on the press, and the first is the communist one; a press
should ensure the stability of a society by educating and empowering people,
although, in practice, the press in past communist countries turned out to be a
machinery for propaganda. The second model is an authoritarian press; a press
controlled by authoritarian rulers and ran by selected journalists. The third
model is the polar opposite; a press that is free from any governmental control
akin to the libertarian idea of the ``marketplace of ideas.'' Besides the free
exchange of ideas, the theory posits media as a watchdog for the government and
ruling elites. The fourth model follows the libertarian ideals but takes also
social responsibility into account with aspects such as enlightenment,
education, and meditation of societal conflicts. Although none of the models
have existed in their pure form, the current Western media systems are located
somewhere between the libertarian and social responsibility models. As is
typical to classification systems, the four models were later altered and
fine-tuned. For instance, the Western media systems have been further broken
down to liberal tradition followed in the English-speaking countries, a
corporatist system typical to the Nordic countries, and a polarized pluralist
system typical to Central Europe~\cite{Hallin04}. Nevertheless, all of these
systems share same universal ideals, among these the freedom of expression, the
power and duty to criticize rulers, and many other related ideals.

Against this backdrop, it becomes understandable why digital divide is also a
political concept; unstable governments, corruption, lack of transparency and
accountability, security and secrecy, and related factors are a reality in many
developing countries~\cite{Thorpe84}. Such conditions not only hamper
journalism, whether offline or online, but also undermine the more or less
universal ideals about media.

These ideals are also taken to reflect the ``maturity'' of online
media. A mature online media should not be only about yellow, paparazzi, or
gonzo journalism---even though the digital environment and its engagement
dynamics tend to perhaps push journalism into this direction. As with the four
theories, the normative stance should be acknowledged; taking something like
\textit{The Intercept} as an ideal type is not universally shared even among
journalists in the Western countries. That said, some of the ideals---such as
the function of a press as a watchdog of ruling elites---are shared also by
journalists in many developing countries, although ``it is not enough to sail in
a turbulent sea with a small boat'' \cite[p.~31]{Khan14}. There is also a more
concrete reason behind this normative maturity notion. Running a media needs
money, and money from Western countries---whether through investments,
developing aid, or something else---tends to come with the same idealistic
strings attached. Some of these strings, such as the freedom of information (but
not necessarily expression), have been endorsed also by the United Nations and
related organizations since the 1970s onward~\cite{Rooy78}. Particularly during
the heydays of the Cold War, mainstream Western media was typically seen as an
essential building block in the development of a well-functioning civil society
and thus democracy~\cite{Miller09}. While the strings may have loosened, they
are still there, partly due to free trade and global regulatory pressure for
national media systems~\cite{Puppis08}. The strings lead to the economic
perspective in turn.

On the supply side of economy, a robust infrastructure is required for online
media to emerge. While states may supply infrastructures for online media and
dictate its content in communist and authoritarian countries, economic
incentives and entrepreneurial ambition are required to establish new online
media in other countries. In this regard, limited capital, lack of skilled
labor, and low tolerance of financial risks have been a long-standing problem
particularly for small and medium-sized enterprises in developing
countries~\cite{Anderson11}. Yet infrastructure can be also understood broadly
to cover not only economic and technological aspects but also socioeconomic
factors, such as education and basic income above extreme poverty.

Indeed, existing studies indicate that income and education are among the most
important aspects distinguishing the use of digital technologies in developed
and developing countries~\cite{Karar19}. These aspects nurture the demand
side. If people live in poverty, they are unlikely to pay for online
media. Conversely: when people have money and are willing to spend it on online
media, there is also a demand for online content. On this demand side, there is
not necessarily a notable gap between developed and developing
countries---getting people to pay for online content is a problem
everywhere. This problem is further associated with the iron grip of
multinational technology companies who, too, have not stood idle; in many
developing countries, the Internet and Facebook are
synonyms~\cite{Shearlaw16}. Besides all other problems and obstacles, the Big
Tech entanglement does not at least make things easier for those attempting to
build online media outlets in developing countries.

\section{Research Design}\label{sec: research design}

\subsection{Data}

\begin{figure}[th!b]
\centering
\includegraphics[width=\linewidth, height=4cm]{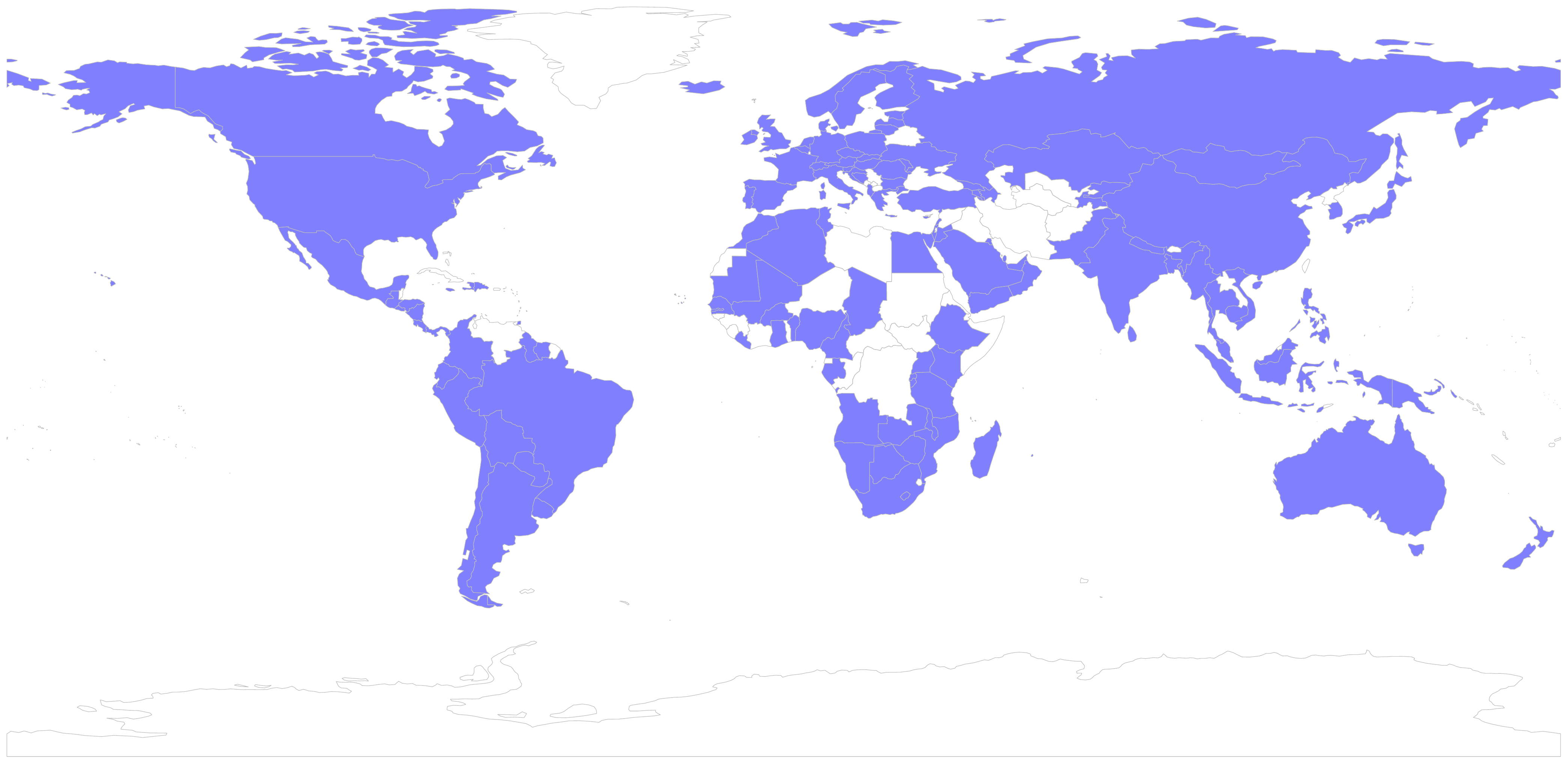}
\caption{Countries Covered in the Dataset}
\label{fig: countries}
\end{figure}

The dataset assembled covers $n = 134$ countries and $t = 10$ years, spanning a
period between 2007 and 2016. The assembling was done with the goal of keeping
the amount of missing values at minimum. Although some countries and years had
to be thus dropped, the coverage across the world is good, as can be concluded
from Fig.~\ref{fig: countries}.

Two data sources were used: the World Bank's online data portals and the
varieties of democracy (\text{V-Dem}) project. The latter source has been widely
used in social sciences~(see \cite{Lewczuk21, Ruohonen21TFSC}, among others), and for a good reason; currently, the \text{V-Dem} project provides the
most comprehensive comparative dataset on democracy, civil liberties, digital
society, and related phenomena. The data is supposedly also robust, as
exemplified the lack of missing values, which, unfortunately, are a problem with
the World Bank data even after deleting many country-year pairs. Following
recent research~\cite{Ruohonen21EJC}, these missing values were interpolated
with cubic~splines~\cite{akima}.

\subsection{Dependent Variables}

\begin{figure}[th!b]
\centering
\includegraphics[width=\linewidth, height=10cm]{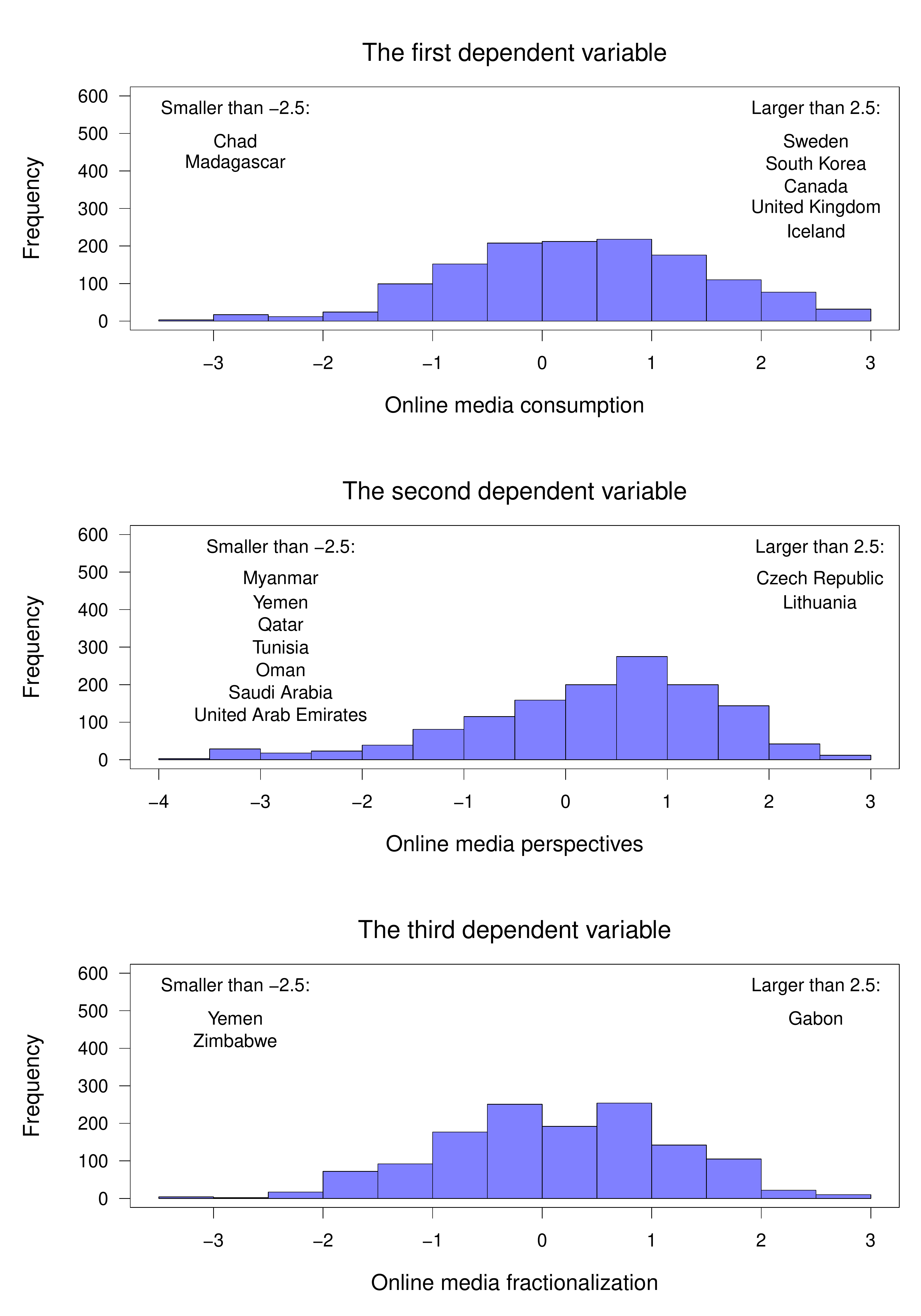}
\caption{Presence of Online Media}
\label{fig: omedia}
\end{figure}

Three dependent variables are used \citep[pp.~325--326]{VDem21b}. The first is
the consumption of online media by domestic audiences. In the \text{V-Dem}
dataset this variable is measured with a four-item Likert-scale: (0)~no one
consumes domestic online media; (1)~limited consumption of domestic online
media; (2)~relatively common consumption of domestic online media; and
(3)~almost everyone consumes domestic online media. Like many variables in the
dataset, it is converted to an ordinal scale that resembles z-values; the value
zero approximates the mean of all country-year pairs of the variable in the
\text{V-Dem} dataset~\citep[pp.~24-25]{VDem21c}. With this scaling, indeed, the
variable's empirical distribution resembles the normal distribution even after
the necessary removal of countries, as can be concluded from the topmost plot in
Fig.~\ref{fig: omedia}. Negative values indicate infrequent consumption of
domestic online media, whereas large positive values correspond with a
well-consumed online domestic media. In the sample used the most consumed
domestic online media are located in Sweden, South Korea, Canada, the United
Kingdom, and Iceland.

The second dependent variable measures the diversity of political perspectives
expressed in online media. Again, a four-item Likert scale is used (before
scaling): (0)~only a government's perspectives are represented; (1)~also
perspectives of semi-official, government-approved opposition are represented;
(3)~there is at least one major online outlet representing all perspectives
important to a society; (4)~there are many outlets representing a wide array of
perspectives. When looking at the middle plot in Fig.~\ref{fig: omedia}, the
interpretation is fairly straightforward with respect to non-democratic countries.

The third and final dependent variable asks a question about whether major
domestic online outlets represents a major political event similarly. Before
scaling, this question is yet again answered with a four-item Likert scale:
major outlets (1)~give opposing views; (2)~differ greatly in their reporting of
major political events; (3)~sometimes agree; (4)~mostly agree; and (5)~there are
only small differences in reporting of big political events. Thus, the variable
measures political consensus of online media; the higher a value, the higher the
consensus. Alternatively: because major newspapers are state-controlled in many
authoritarian countries, reporting mainly in line of political establishments,
more divergence may be found from online outlets. When looking at the countries
with low and high scaled scores in Fig.~\ref{fig: omedia}, either interpretation
seems plausible.

The three variables are assumed to represent different viewpoints on the
maturity of an online media ecosystem in a given country. When online media is
well-established and widely consumed, it represents also a plurality of
different political viewpoints, and so forth. Statistically, however, the
variables are unrelated; the maximum correlation coefficient between the three
variables is only $0.33$. This lack of forceful statistical dependence justifies
modeling of the three variables separately even though they are understood to be
parts of the same underlying online media maturity concept.

\subsection{Independent Variables}\label{sec: independent variables}

The independent variables are enumerated in Table~\ref{tab: variables} within
which the third column refers to the online source from which a given dataset
was retrieved. That is to say, the primary source may differ for some of the
World Bank's datasets. For instance, the International Telecommunication Union
is the primary data source for the fixed broadband and mobile cellular
subscription amounts. The grouping of the variables aligns with the analytical
framework in Fig.~\ref{fig: framework}; each group is represented by four
carefully selected variables.

\begin{table*}[th!b]
\centering
\caption{Independent Variables (refer to Fig.~\ref{fig: framework} for the
  analytical motivation)}
\label{tab: variables}
\begin{tabular}{lllr}
\toprule
Variable & Category & Description & Retrieval source \\
\hline
GDP & Economy & GDP per capita (current US\$) & \cite{WorldBank21b} \\
FDI & Economy & Foreign direct investment inward flows (\% GDP) & \cite{WorldBank21c} \\
Procurement & Economy & Government procurement of advanced technology products & \cite{WorldBank21e} \\
State economy & Economy & State control of important sectors of economy & \cite{VDem21a} \\
Electricity & Infrastructure & Electricity and telephony infrastructure & \cite{WorldBank21e} \\
Education & Infrastructure & Quality of the education system & \cite{WorldBank21e} \\
Broadband & Infrastructure & Fixed broadband subscriptions & \cite{WorldBank21f} \\
Mobile & Infrastructure & Mobile cellular subscriptions & \cite{WorldBank21g} \\
Rule of law & Administration & Enforcement and compliance with laws & \cite{VDem21a} \\
Accountability & Administration & Accountability of a government & \cite{VDem21a} \\
Transparency & Administration & Transparency of a government's policy-making & \cite{WorldBank21e} \\
Corruption & Administration & Ethics and corruption & \cite{WorldBank21e} \\
Liberal democracy & Politics & Extent of liberal democracy & \cite{VDem21a} \\
Freedom of expression & Politics & Freedom of expression of media and people & \cite{VDem21a} \\
Political arrests & Politics & Arrests due to political online content & \cite{VDem21a} \\
Content removal & Politics & Government removal of online content & \cite{VDem21a} \\
\bottomrule
\end{tabular}
\end{table*}

The economic and infrastructure variables have been widely used in previous
research~\cite{Pick07, Pick15}. It is therefore more relevant to make few
remarks about the operationalization of the variables. The variables from the
World Bank's TCdata360 portal \cite{WorldBank21e}, such as those for corruption
or government transparency, are scalar-valued indices whose values range from
one to seven; the higher a value, the better the outcome; a value seven implies
an unlikely case that no corruption whatsoever is present in a country, for
instance.

The political variables from the \text{V-Dem} project \cite{VDem21b} require a
more thorough elaboration. The variable for liberal democracy is a composite
index, constructed from four individual variables. It measures the protection of
minority groups and individuals against the tyranny of the state and the
majority; higher values indicate more liberalism. The freedom of expression
variable, too, is a composite index measuring press freedom, academic and
cultural expression, as well as the freedom of people to discuss political
issues at home and in public spaces. Political arrest measures the likelihood
that a citizen will be arrested upon posting political content
online. A~four-item Likert-scale is used: an arrest is (0)~extremely likely, (1)
likely, (2) unlikely, or (3)~extremely unlikely. The removal of online content
by a government, in turn, is measured with a five-item scale: (0)~any content
can be removed at will; (1)~law protects only politically uncontroversial
issues; (2)~legal ambiguity exists for removal; (3)~political speech is mostly
protected yet politically controversial content may be removed; (4)~a law
protects all political speech such that a government can only remove content
according to rigorous legal criteria. By hypothesis, at least one of these
political variables should be associated either with the perspectives dependent variable or the fractionalization dependent variable, or~both.

\subsection{Methods}

The dataset's nature is time series cross-sectional data, but, since $n > t$ to
a fair margin, panel data is a better characterization. The classical estimator
for such data is ordinary least squares augmented with dummy variables for the
years and countries. In addition to problems with accuracy~\cite{Nickell81}, as
manifested in the covariance structure and thus standard errors of the
regression coefficients, the least squares estimator ignores all longitudinal
dynamics; autocorrelation in the error term is thus another problem. Adding a
lagged dependent variable is an option, but it leads to additional problems.

To address these issues, the dynamic Arellano-Bond panel data estimator
\cite{Arellano91} is used, as implemented in an R
package~\cite{plm}. Essentially, the estimator uses a lagged dependent variable
and first differences to get rid of the constant ($\alpha$) and the
cross-sectional effects ($c_i$), such that in a toy model with only one
independent variable, an equation $y_{it} = \alpha + \beta_1 y_{it-1} + \beta_2
x_{it} + c_i + \epsilon_{it}$ becomes $\Delta y_{it} = \beta_1 \Delta y_{it-1} +
\beta_2\Delta x_{it} + \Delta \epsilon_{it}$. This, however, creates a moving
average process to the error term, which the estimator accounts for with
instrumental variables. For the models estimated, the second, third, and fourth
lags are used as instruments for the lagged dependent variable as well as all
independent variables. In addition, lagged differences, $\Delta x_{it-1}$, are
added for the independent variables.

As for basic diagnostic checks, first-order autocorrelation (AR) should be
present but second-order should be absent; a test developed by Arellano and
Bond~\cite{Arellano91} is provided by the implementation used. And as for
practical estimation, a $\ln(x + 1)$ is transformation is applied to all
independent variables that do not attain negative values. Finally, nine
additional dummy variables are included for annual effects, although these are
not reported for brevity.

\section{Results}\label{sec: results}

The regression results are summarized in Table~\ref{tab: results}, which shows
three models corresponding with the three dependent variables. The basic
diagnostic checks hold for all models; the null hypotheses of no second-order
autocorrelation hold. Wald's tests indicate that the coefficients are jointly
statistically significant in all models. Thus, to proceed, the models can be
briefly disseminated as follows:

\begin{table*}[ht]
\centering
\caption{Regression Results}
\label{tab: results}
\begin{tabular}{llll}
 \hline
 & \multicolumn{3}{c}{Dependent variable $(y_{it})$} \\
 \cmidrule{2-4}
& Consumption & Perspectives & Fractionalization \\
  \hline
$\Delta y_{it-1}$ & $\phantom{-}0.578^{***}$ & $\phantom{-}0.259^{***}$ & $\phantom{-}0.560^{***}$ \\
 \cmidrule{2-4}
  $\Delta\ln(\textmd{GDP} + 1)_{it}$ & $-0.119^{\phantom{***}}$ & $-0.057^{\phantom{***}}$ & $\phantom{-}0.038^{\phantom{***}}$ \\
  $\Delta\ln(\textmd{GDP} + 1)_{it-1}$ & $\phantom{-}0.343^{**}\phantom{*}$ & $\phantom{-}0.102^{\phantom{***}}$ & $-0.071^{\phantom{***}}$ \\
  $\Delta\textmd{(FDI)}_{it}$ & $<0.001^{\phantom{**}}$ & $< 0.001^{\phantom{***}}$ & $< 0.001^{\phantom{***}}$ \\
  $\Delta\textmd{(FDI)}_{it-1}$ & $\phantom{-}0.002^{**}\phantom{*}$ & $< 0.001^{\phantom{***}}$ & $< 0.001^{\phantom{***}}$ \\
  $\Delta\ln(\textmd{Procurement} + 1)_{it}$ & $-0.001^{\phantom{***}}$ & $\phantom{-}0.017^{\phantom{***}}$ & $-0.023^{\phantom{***}}$ \\
 $\Delta\ln(\textmd{Procurement} + 1)_{it-1}$ & $\phantom{-}0.039^{\phantom{***}}$ & $-0.005^{\phantom{***}}$ & $-0.010^{\phantom{***}}$ \\
  $\Delta\textmd{(State economy)}_{it}$ & $-0.085^{\phantom{***}}$ & $\phantom{-}0.064^{\phantom{***}}$ & $\phantom{-}0.033^{\phantom{***}}$ \\
 $\Delta\textmd{(State economy)}_{it-1}$ & $\phantom{-}0.069^{\phantom{***}}$ & $\phantom{-}0.140^{\phantom{***}}$ & $-0.095^{\phantom{***}}$ \\
\cmidrule{2-4}
  $\Delta\ln(\textmd{Electricity} + 1)_{it}$ & $\phantom{-}0.096^{\phantom{***}}$ & $-0.011^{\phantom{***}}$ & $\phantom{-}0.011^{\phantom{***}}$ \\
  $\Delta\ln(\textmd{Electricity} + 1)_{it-1}$ & $-0.047^{\phantom{***}}$ & $-0.036^{\phantom{***}}$ & $\phantom{-}0.017^{\phantom{***}}$ \\
  $\Delta\ln(\textmd{Education} + 1)_{it}$ & $\phantom{-}0.001^{\phantom{***}}$ & $\phantom{-}0.098^{\phantom{***}}$ & $\phantom{-}0.024^{\phantom{***}}$ \\
  $\Delta\ln(\textmd{Education} + 1)_{it-1}$ & $\phantom{-}0.035^{\phantom{***}}$ & $\phantom{-}0.022^{\phantom{***}}$ & $\phantom{-}0.031^{**}\phantom{*}$ \\
  $\Delta\ln(\textmd{Broadband} + 1)_{it}$ & $-0.018^{\phantom{***}}$ & $\phantom{-}0.009^{\phantom{***}}$ & $\phantom{-}0.003^{\phantom{***}}$ \\
  $\Delta\ln(\textmd{Broadband} + 1)_{it-1}$ & $\phantom{-}0.011^{\phantom{***}}$ & $\phantom{-}0.002^{\phantom{***}}$ & $-0.032^{***}$ \\
  $\Delta\ln(\textmd{Mobile} + 1)_{it}$ & $-0.288^{\phantom{***}}$ & $\phantom{-}0.030^{\phantom{***}}$ & $-0.024^{\phantom{***}}$ \\
 $\Delta\ln(\textmd{Mobile} + 1)_{it-1}$ & $\phantom{-}0.120^{\phantom{***}}$ & $\phantom{-}0.010^{\phantom{***}}$ & $\phantom{-}0.087^{\phantom{***}}$ \\
\cmidrule{2-4}
  $\Delta\ln(\textmd{Rule of law} + 1)_{it}$ & $\phantom{-}1.098^{\phantom{***}}$ & $\phantom{-}1.386^{\phantom{***}}$ & $\phantom{-}0.470^{\phantom{***}}$ \\
  $\Delta\ln(\textmd{Rule of law} + 1)_{it-1}$ & $-0.984^{\phantom{***}}$ & $-1.045^{\phantom{***}}$ & $\phantom{-}0.329^{\phantom{***}}$ \\
  $\Delta\ln(\textmd{Accountability} + 1)_{it}$ & $\phantom{-}0.060^{\phantom{***}}$ & $-0.021^{\phantom{***}}$ & $-0.017^{\phantom{***}}$ \\
  $\Delta\ln(\textmd{Accountability} + 1)_{it-1}$ & $-0.051^{\phantom{***}}$ & $\phantom{-}0.101^{\phantom{***}}$ & $-0.020^{\phantom{***}}$ \\
  $\Delta\ln(\textmd{Transparency} + 1)_{it}$ & $\phantom{-}0.530^{**}\phantom{*}$ & $\phantom{-}0.157^{\phantom{***}}$ & $-0.437^{\phantom{***}}$ \\
  $\Delta\ln(\textmd{Transparency} + 1)_{it-1}$ & $-0.370^{\phantom{***}}$ & $-0.153^{\phantom{***}}$ & $\phantom{-}0.273^{\phantom{***}}$ \\
  $\Delta\ln(\textmd{Corruption} + 1)_{it}$ & $-0.044^{\phantom{***}}$ & $-0.168^{\phantom{***}}$ & $\phantom{-}0.078^{\phantom{***}}$ \\
$\Delta\ln(\textmd{Corruption} + 1)_{it-1}$ & $-0.013^{\phantom{***}}$ & $\phantom{-}0.062^{\phantom{***}}$ & $-0.139^{\phantom{***}}$ \\
\cmidrule{2-4}
  $\Delta\ln(\textmd{Liberal democracy} + 1)_{it}$ & $\phantom{-}0.102^{\phantom{***}}$ & $-0.412^{\phantom{***}}$ & $-0.739^{\phantom{***}}$ \\
  $\Delta\ln(\textmd{Liberal democracy} + 1)_{it-1}$ & $-0.446^{\phantom{***}}$ & $\phantom{-}0.007^{\phantom{***}}$ & $-0.186^{\phantom{***}}$ \\
  $\Delta\ln(\textmd{Freedom of expression} + 1)_{it}$ & $\phantom{-}0.206^{\phantom{***}}$ & $\phantom{-}0.153^{\phantom{***}}$ & $-0.058^{\phantom{***}}$ \\
  $\Delta\ln(\textmd{Freedom of expression} + 1)_{it-1}$ & $-0.011^{\phantom{***}}$ & $-0.753^{\phantom{***}}$ & $\phantom{-}0.344^{\phantom{***}}$ \\
  $\Delta\textmd{(Political arrests)}_{it}$ & $-0.002^{\phantom{***}}$ & $\phantom{-}0.377^{***}$ & $\phantom{-}0.049^{\phantom{***}}$ \\
  $\Delta\textmd{(Political arrests)}_{it-1}$ & $\phantom{-}0.002^{\phantom{***}}$ & $\phantom{-}0.017^{\phantom{***}}$ & $-0.062^{\phantom{***}}$ \\
  $\Delta\ln(\textmd{Content removal} + 1)_{it}$ & $-0.059^{\phantom{***}}$ & $\phantom{-}0.595^{***}$ & $\phantom{-}0.519^{***}$ \\
  $\Delta\ln(\textmd{Content removal} + 1)_{it-1}$ & $\phantom{-}0.158^{\phantom{***}}$ & $-0.060^{\phantom{***}}$ & $-0.227^{\phantom{***}}$ \\
\hline
$\textmd{H}_0$: no AR(1), $p$-values & $< 0.001^{\phantom{***}}$ & $< 0.001^{\phantom{***}}$ & $< 0.001^{\phantom{***}}$ \\
$\textmd{H}_0$: no AR(2), $p$-values & $\phantom{-}0.334^{\phantom{***}}$ & $\phantom{-}0.564^{\phantom{***}}$ & $\phantom{-}0.765^{\phantom{***}}$ \\
\hline
\end{tabular}
\end{table*}

\begin{enumerate}
\itemsep 3pt
\item{Three coefficients are statistically significant in the first model. These
  are: GDP per capita, inward FDI flows, and the transparency of governments'
  policy-making processes. The signs of the corresponding coefficients are also
  positive, as was expected. Though, the magnitude of the coefficient for the
  FDI variable is too small to warrant specific attention. When people have more
  money, the demand for online media increases, and when a government is
  transparent, there is more for journalists to report and readers to read.}
\item{Two coefficients are statistically significant in the second model:
  political arrests and governments' removal of online content. When recalling
  the operationalization (see Section~\ref{sec: independent variables}), the
  positive signs for the coefficients are also as expected. When there is no
  fear of being arrested for reporting multiple viewpoints, the plurality of
  perspectives expressed in online media increases. Analogous reasoning applies
  when a government rarely removes online content.}
\item{Three coefficients are statistically significant in the third model. The
  first is the one for education quality, which has a positive sign, which seems
  reasonable. The second one for broadband subscriptions has a negative sign,
  however, which seems illogical or at least unexpected. The third is again for
  the coefficient of the content removal variable. Thus: when removal of
  political content by a government is unlikely due to legal protections, there
  are also less disagreements in reporting major political events in online
  media.}
\end{enumerate}

These results align with prior expectations and seem sensible in general,
excluding the one negative sign noted. Although statistical significance is
hardly the only (or even the right) way to proceed with statistical inference,
what may seem a little surprising is the lack of statistically significant
effects for many of the variables. For instance, the other coefficients for the
infrastructure variables are not statistically significant even at the
conventional $p < 0.05$ level. Alternative model specifications do not change
the situation. A~potential explanation for the lack of statistically significant
effects for the infrastructure variables relates to the noted fact that the
infrastructure already exists in many countries. When considering the second and
third models, the perspectives and fractionalization, it may also be that the
infrastructures are already \textit{too} good in some countries with communist
or authoritarian media systems. In other words, the surveillance capabilities in
many countries may well establish a certain deterrence
effect~\cite{Ruohonen21TFSC}, leading to self-censorship that may curtail press
freedom and thus decrease the plurality of political viewpoints by journalists
in online media. Besides these speculations, there is a potential statistical
explanation.

It should be also recalled that the models estimated are dynamic panel data
models, and the magnitudes of the lagged dependent variables indicate a high
degree of persistence. A general lack of longitudinal variance has been observed
also previously with the \text{V-Dem} data~\cite{Ruohonen21EJC}. Furthermore,
Wald's $\chi^2$-tests indicate that the dummy variables for annual effects are
not jointly statistically significant in the second and third models. These
points can be used to argue that the situation has remained mostly unchanged in
the 2007--2016 period observed. Nevertheless, to interpret the results against
the framework in Fig.~\ref{fig: framework}, there is enough evidence to conclude
that economy, state administration, and politics (but not necessarily
infrastructure) contribute to the maturity of online~media.

\section{Conclusion}\label{sec: conclusion}

This paper examined the maturity of online media in time series cross-sectional
framework covering 134 countries in a period spanning ten years. Maturity was
defined according to consumption of online media, political perspectives
expressed in online media, and fractionalization of online media in terms of
reporting major political events. A simple four-dimensional analytical model
(economics, infrastructure, administration, and politics) was used to motivate
the empirical analysis. Based on dynamic panel data estimator, at least one
variable belonging to each dimension was statistically significant in explaining
the three concepts of maturity; namely, GDP per capita, inward FDI flows,
transparency of governmental administration, potential arrests due to political
online content, content removal by a government, education, and broadband
subscriptions. All in all, the results align with the prior theoretical
speculations.

But, by design, the time series cross-sectional research setup can only touch
the surface of the question of online media maturity. While sensible for the
setup, particularly the independent variables only indirectly proxy the
underlying analytical dimensions in Fig.~\ref{fig: framework}. To gain a more
realistic perspective, survey methodology would be more suitable in order to
gain explicit answers about the challenges and obstacles for building new
digital media particularly in developing countries. But then again, obtaining
longitudinal data with a global scope from surveys is an extremely difficult
task without substantial resources. Given this backdrop, it seems sensible to
recommend that further work should be done to quantify more variables in
existing time series cross-sectional data frameworks such as \text{V-Dem}. For
instance, public service media has been observed to be an important factor
explaining the diversity of perspectives expressed in national media
systems~\citep{Humprecht17}, and quantifying the existence of public
broadcasting should be a relatively easy task.

Public broadcasting and public service media in general are good examples also
from another perspective. Government intervention has been argued to be
necessary to bridge a digital divide in many developing
countries~\cite{Ogondo16}. On one hand, this point can be taken to reflect
critically on the ideals of libertarian media systems in English-speaking
countries. It may be that economic opportunities and civil liberties are not
enough for a new online media system to emerge; the corporatist model in the
Nordic countries may be a viable option. On the other hand, particularly in
developing countries the dangers of state control are real; having too much
government involvement may turn toward the communist and authoritarian media
system types via which the freedom of expression and other civil liberties may
be endangered. Every coin has two sides. Finally, it is worth noting that a decade may be too short---after all, the evolution of the printing press toward maturity took centuries.

\begin{acks}
This paper was partially funded by the Strategic Research Council at the Academy
of Finland (grant number~327391).
\end{acks}

\bibliographystyle{abbrv}

\end{document}